\documentclass[uplatex,review]{elsarticle}
\usepackage{hyperref}
\usepackage{amssymb}
\usepackage{graphicx}
\usepackage{bm}
\usepackage{braket}
 
\journal{Physics Letters B}

\begin{document}

\begin{frontmatter}

\title{
``$\mathbf{{\textit K^-}{\textit p}{\textit p}}$'', a ${\overline{K}}$-Meson Nuclear Bound State, Observed in $^{3}{\rm He}({K^-}, {\Lambda} p)n$ Reactions
}

\author{J-PARC E15 collaboration}
\author[OSAKA]{S.~Ajimura}
\author[RIKEN]{H.~Asano}
\author[Victoria]{G.~Beer}
\author[SMI]{C.~Berucci}
\author[Seoul]{H.~Bhang}
\author[INFNHH]{M.~Bragadireanu}
\author[SMI]{P.~Buehler}
\author[Torino,TorinoU]{L.~Busso}
\author[SMI]{M.~Cargnelli}
\author[Seoul]{S.~Choi}
\author[Frascati]{C.~Curceanu}
\author[KEK]{S.~Enomoto}
\author[TITEC]{H.~Fujioka}
\author[Tokyo]{Y.~Fujiwara}
\author[OsakaE]{T.~Fukuda}
\author[Frascati]{C.~Guaraldo}
\author[JAEA]{T.~Hashimoto}
\author[Tokyo]{R.~S.~Hayano}
\author[OSAKA]{T.~Hiraiwa}
\author[KEK]{M.~Iio}
\author[Frascati]{M.~Iliescu}
\author[OSAKA]{K.~Inoue}
\author[Kyoto]{Y.~Ishiguro}
\author[Tokyo]{T.~Ishikawa}
\author[KEK]{S.~Ishimoto}
\author[RIKEN]{K.~Itahashi}
\author[RIKEN,TITEC]{M.~Iwasaki\corref{mycorrespondingauthor}}
\ead{masa@riken.jp}
\author[Tokyo]{K.~Kanno}
\author[Kyoto]{K.~Kato} 
\author[RIKEN]{Y.~Kato}
\author[OSAKA]{S.~Kawasaki}
\author[TUM]{P.~Kienle$^{\dagger,}$}
\author[TITEC]{H.~Kou}
\author[RIKEN]{Y.~Ma}
\author[SMI]{J.~Marton}
\author[Tokyo]{Y.~Matsuda}
\author[OsakaE]{Y.~Mizoi}
\author[Torino]{O.~Morra}
\author[Kyoto]{T.~Nagae}
\author[OSAKA]{H.~Noumi}
\author[TohokuELPH,RIKEN]{H.~Ohnishi}
\author[RIKEN]{S.~Okada}
\author[RIKEN]{H.~Outa}
\author[Frascati]{K.~Piscicchia}
\author[OSAKA]{Y.~Sada}
\author[OSAKA]{A.~Sakaguchi}
\author[RIKEN]{F.~Sakuma\corref{mycorrespondingauthor}}
\ead{sakuma@ribf.riken.jp}
\author[KEK]{M.~Sato}
\author[Frascati]{A.~Scordo}
\author[KEK]{M.~Sekimoto}
\author[Frascati]{H.~Shi}
\author[OSAKA]{K.~Shirotori}
\author[Frascati,INFNHH]{D.~Sirghi}
\author[Frascati,INFNHH]{F.~Sirghi}
\author[SMI]{K.~Suzuki}
\author[KEK]{S.~Suzuki}
\author[Tokyo]{T.~Suzuki}
\author[JAEA]{K.~Tanida}
\author[Sweden]{H.~Tatsuno}
\author[TITEC]{M.~Tokuda}
\author[OSAKA]{D.~Tomono}
\author[KEK]{A.~Toyoda}
\author[TohokuELPH]{K.~Tsukada}
\author[Frascati,TUM]{O.~Vazquez~Doce}
\author[SMI]{E.~Widmann}
\author[RIKEN,OSAKA]{T.~Yamaga\corref{mycorrespondingauthor}}
\ead{takumi.yamaga@riken.jp}
\author[Tokyo,RIKEN]{T.~Yamazaki}
\author[RIKEN]{Q.~Zhang}
\author[SMI]{J.~Zmeskal}

\address[OSAKA]{Osaka University, Osaka, 567-0047, Japan}
\address[RIKEN]{RIKEN, Wako, 351-0198, Japan}
\address[Victoria]{University of Victoria, Victoria BC V8W 3P6, Canada }
\address[SMI]{Stefan-Meyer-Institut f\"{u}r subatomare Physik, A-1090 Vienna, Austria }
\address[Seoul]{Seoul National University, Seoul, 151-742, South Korea }
\address[INFNHH]{National Institute of Physics and Nuclear Engineering - IFIN HH, Bucharest - Magurele, Romania }
\address[Torino]{INFN Sezione di Torino, 10125 Torino, Italy }
\address[TorinoU]{Universita' di Torino, Torino, Italy }
\address[Frascati]{Laboratori Nazionali di Frascati dell' INFN, I-00044 Frascati, Italy }
\address[KEK]{High Energy Accelerator Research Organization (KEK), Tsukuba, 305-0801, Japan }
\address[TITEC]{Tokyo Institute of Technology, Tokyo, 152-8551, Japan }
\address[Tokyo]{The University of Tokyo, Tokyo, 113-0033, Japan }
\address[OsakaE]{Osaka Electro-Communication University, Osaka, 572-8530, Japan }
\address[JAEA]{Japan Atomic Energy Agency, Ibaraki 319-1195, Japan }
\address[Kyoto]{Kyoto University, Kyoto, 606-8502, Japan }
\address[TUM]{Technische Universit\"{a}t M\"{u}nchen, D-85748, Garching, Germany }
\address[TohokuELPH]{Tohoku University, Sendai, 982-0826, Japan }
\address[Sweden]{Lund University, Lund, 221 00, Sweden }

\cortext[mycorrespondingauthor]{Corresponding authors}

\begin{abstract}
We observed a distinct peak in the $\Lambda p$ invariant mass spectrum of  $^{3}{\rm He}(K^-, \, \Lambda p)n$, well below the mass threshold of $m_K + 2 m_p$. 
By selecting a relatively large momentum-transfer region $q = 350 \sim 650$ MeV/$c$, one can kinematically separate the peak from the quasi-free process, $\overline{K}N \rightarrow \overline{K}N$ followed by the non-resonant absorption by the two spectator-nucleons $\overline{K}NN \rightarrow \Lambda N$.
We found that the simplest fit to the observed peak gives us a Breit-Wigner pole position at $B_{\rm {\it Kpp}} = 47 \pm 3 \,  (stat.) \,^{+3}_{-6} \,(sys.)$ MeV having a width $\Gamma_{\rm {\it Kpp}} = 115 \pm 7 \,  (stat.) \,^{+10}_{-20} \,(sys.)$ MeV, and the $S$-wave Gaussian reaction form-factor parameter $Q_{\rm {\it Kpp}} = 381 \pm 14 \, (stat.)\,^{+57}_{-0} \,(sys.)$ MeV/$c$, as a new form of  the nuclear bound system with strangeness -- ``$K^-pp$''.
\end{abstract}


\end{frontmatter}



\section{Introduction}
   Since the prediction of the $\pi$-meson by Yukawa \cite{Yukawa35}, the long-standing question has been whether a mesonic nuclear bound state exists,
   {\it{i.e.}}, whether a meson forms {\it a quantum state} at an eigen-energy $E_{M}$ below the intrinsic mass $m$ without promptly vanishing in nuclear media.
   If it exists,  it means that a meson ($\overline{q}q$) forms a quantum state where baryons ($qqq$) exist as nuclear medium.
 There are many important subjects to study, {\it e.g.}, how hadron masses are generated from $\sim$\,massless particles: quarks ($m_q \sim$ few MeV/$c^2$) and gluons ($m_g=0$), how the properties of these mesons change in the nuclear medium, how hadrons are confined in the nuclear media, and the equation-of-state in nuclear (or star) matter.
   Therefore, many mesons have been examined over the past century, to see whether  a mesonic nuclear bound state exists below the mass threshold with a binding energy $B_{M} \equiv m - E_{M}$, but there has been no clear evidence for their existence.

   The $\pi N$ $S$-wave interaction is repulsive, so there is no nuclear bound state much deeper than the atomic states \cite{Geissel02}.
   What about the second-lightest meson with an $s$-quark, the kaon?
   After the long standing {\it{``kaonic hydrogen puzzle'' }} was resolved \cite{Iwasaki97,Beer,Bazzi}, the strong ${\overline K}N$ attractive interaction was established in the isospin $I=0$ channel.  
   This leads us naturally to the ansatz that the $\Lambda(1405)$ could be a $K^- p$ nuclear bound state, rather than a three-quark $\Lambda$-baryonic-state as it is named, {\it{i.e.}}, the name implies that it is a first excited state of the $\Lambda$ baryon whose excitation is caused by the constituent-quark internal-motion. 
   A recent lattice QCD calculation also supports the $K^- p$ picture \cite{Hall15}.
   Akaishi-Yamazaki predicted the existence of kaonic nuclear bound states assuming the $\Lambda(1405)$ be a $K^- p$ bound state \cite{AY1}.
   The simplest predicted kaonic nuclear system, $\overline{K}NN$ symbolically denoted as ``$K^- pp$'', has charge $+1$, $I=\frac{1}{2}$ and $J^P = 0^-$, with a binding energy $B_{\rm {\it Kpp}}$ = 48 MeV (measured from $M(Kpp) \equiv m_K + 2 m_p \approx$ 2370 MeV/$c^2$) and a partial mesonic decay width  $\Gamma_{\pi Y \!N} $ = 61 MeV \cite{AY2}.

   Triggered by this prediction, many studies were undertaken.
   Theoretically, the existence of the kaonic bound states is well supported, but the results are widely scattered: binding energies ($B_{\rm {\it Kpp}} \approx 10 \sim 100$ MeV) and partial mesonic decay widths ($\Gamma_{\pi Y \!N} \approx 40 \sim 100$ MeV), {\it e.g.}, \cite{Gal,Weise,Oset,Dote}, while the total decay width $\Gamma_{Kpp}$ (including non-mesonic decay channels) is not yet calculated. 
   Experimentally, there have been many searches for ``$K^- pp$'', with reports of possible candidates \cite{finuda,disto,e27} as well as contradictory results \cite{hades,oton}, leaving the matter both controversial and unsettled. 

\section{J-PARC E15 Experiment}

   To search for the ``$K^- pp$'', the most straightforward experiment is the $K\!^- \!+ \!^2$He reaction {\it{below the $M(Kpp)$ mass threshold}}, which is obviously impossible.  
   Instead, we have conducted an experimental search by bombarding a $^3$He target with a 1 GeV/$c$ $K^-$ beam to knock out a nucleon with the kaon, and directly introduce a recoiled virtual-${\overline {K}}$-meson into the residual nucleus.
   At this momentum ($\sqrt{s} \sim 1.8$ GeV for $\overline{K}N$), the single-nucleon elastic-reaction ${\overline {K}}N \rightarrow {\overline {K}}N$ has a very large cross-section, helped by the presence of $Y^*$-resonances ($m_{Y^*} \sim$ 1.8 GeV/$c^2$) \cite{Tanabashi:2018oca}.
  On the other hand, due to the shrinkage of the de Broglie wave-length of the projectile, direct multi-NA,  which produces a severe background in an {\it at-rest-kaon-absorption} experiment to search for ``$K^- pp$'' \cite{oton,Sato}, will be relatively suppressed.
      
      The momentum of the virtual `${\overline {K}}$' is given as $q_{\,\overline {K}} = q_{K n} \equiv |{\bf q}_{K n}|$ ({\it i.\,e.}\,the momentum difference of an incident kaon and the forward neutron ${\bf q}_{K n} \equiv  {\bf p}_{K^-}^{Lab.} - {\bf p}_{n}^{Lab.} $), where the superscript represents that it is in the laboratory-frame, and the single quotation marks represent that it is within the strong interaction range in a nucleus.
      When the `${\overline {K}}$' is backscattered, the  $q_{\,\overline {K}}$ can be as small as $\sim$ 200 MeV/$c$ (the minimum $q$ among the search experiments performed).
    With this condition, a successive reaction between the virtual `${\overline {K}}$' and two `{\it  spectator nucleons}' {\it at-rest in the laboratory-frame} can be efficiently realized.
    This way, a ``$K^- pp$'' can be formed almost at-rest in the laboratory-frame, which makes the formation probability large. 
    In this reaction channel, one can reduce the possible combinations of ``$K^- pp$'' decay particles, because the $s$-quark is conserved in the strong interaction and thus a hadron with an $s$-quark should exist in its decay.
       Thus, one can efficiently conduct invariant mass spectroscopy (decay channel) by having the detector surrounding the target, and missing mass spectroscopy (formation channel) using a forward neutron counter (NC) and a spectrometer to simultaneously detect a forward going neutron (or proton) coming from ${\overline {K}}N \rightarrow {\overline {K}}N$ reaction.
        We designed our apparatus to achieve a mass resolution of $\sigma_M \sim 10$ MeV/$c^2$ both in missing and in invariant mass \cite{Agari}.

   The first-stage experiment, J-PARC E15$^{\rm 1st}$, exhibited a huge peak above $M(Kpp)$ by observing the neutron in NC ($\Delta \theta_{N\!C} \sim 1/20$) \cite{hashimoto}. 
   This spectral peak has a very large cross-section of $\gtrsim$~6 mb/sr in the semi-inclusive quasi-elastic ${\overline{K}}N \rightarrow {\overline{K}}N$ channel at $\theta_n = 0$. 
   Thus, we confirmed that the forward nucleon knockout reaction, ${\overline{K}}N \rightarrow {\overline{K}}N$, is the dominant process at $p_K = 1$ GeV/$c$. 
   It also revealed that there was a large event-excess extending from the quasi-elastic ${\overline{K}}$ bump to the lower mass region.  
   The tail reached to $\sim$ 100 MeV below $M(Kpp)$ ($\sim$ 1 mb/sr).
   However, no significant structure was observed in this tail at any location, where ``$K^- pp$'' candidates were reported \cite{finuda,disto,e27}.

   On the contrary, we found a kinematical anomaly: a peak-like structure was observed in the $\Lambda p$ invariant mass (${\rm{\it IM}}_{\!\Lambda p} \equiv M, \rm{\it{ hereafter}})$ spectrum of the non-mesonic $\Lambda p n$ final state (observed by the $p p \pi^-$-events without requesting an NC hit) below the $M(Kpp)$ mass threshold at low $q_{\Lambda p} ~(= q_{K n} \equiv q, \rm{\it{ hereafter}})$ \cite{sada}.
   This is the simplest final state, which consists of the minimum number of lowest mass baryons without meson emission, so the possible interpretations are limited.
   The most promising interpretation is: the ``$K^- pp$'' is formed by knocking out a neutron, decays to $\Lambda p$, and thus a corresponding peak is seen in the $M$ spectrum.
   
    To significantly improve the statistics of the $\Lambda p n$ final state permitting us to examine this interpretation, we set a higher priority on accumulating events having three charged particle hits around the target {\it without requiring forward neutron detection}.
    The kinematical refit of $p\pi^- p$ (+\,{\it n missing}) to the $\Lambda p n$ final state using energy-momentum conservation was conducted at the analysis stage to prevent biasing the data.
    We succeeded in accumulating 30 times as much data on $p\pi^- p$ events compared to E15$^{1st}$.  

 \begin{figure}[htbp]
 \begin{center}
 \includegraphics[width=10cm]{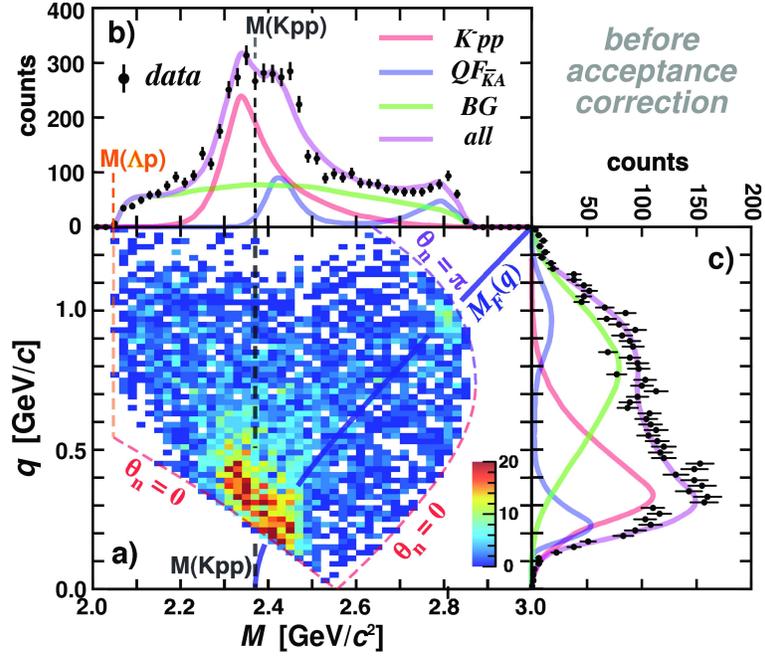}
 \caption{
  \label{fig:data}
$a$) 2D event distribution plot on the $M$ ($=\rm{\it IM}_{\Lambda p}$) and the momentum transfer $q$ ($q_{\Lambda p}$) for the $\Lambda p n$ final state.  
The $M_{F}(q)$ given in Eq.\,\ref{eq:QFA-centroid}, the mass threshold $M(Kpp)$, and the kinematical boundary for $\Lambda p n$ final state, are plotted in the figure. The lower $q$ boundary corresponds to $ \theta_n = 0$ (forward $n$), and the upper boundary corresponds to $\theta_n = \pi$ (backward $n$).
   The histograms of projection onto the $M$ axis $b$),  and onto $q$ axis $c$)  are also given together with the decompositions of the fit result.}
 \end{center}
 \end{figure} 

   The formation channel, $K^-+^3{\rm He} \rightarrow ``K^-pp" + n$, can be uniquely defined by the following two parameters; the $\Lambda p$-invariant mass $M$ and the momentum transfer $q$.
   The event distributions over $M$ and $q$ are given in Fig.\,\ref{fig:data}.
   As shown in the figure, a strong event-concentration observed previously \cite{sada} is confirmed near the mass threshold $M(Kpp)$ at the lower-$q$ side $(M\!c^2, \,qc) \sim (2.37, \,0.25)$ GeV. 

   To our surprise, however, the structure near $M(Kpp)$ cannot be represented as a single Breit-Wigner (B.W.) function, as was na{\"{\i}}vely assumed in the previous paper \cite{sada}.
   Instead, it is more natural to interpret this structure as consisting of at-least two internal substructures originating from different reaction mechanisms. 
   However, the primary reaction $K^-N \rightarrow $`${\overline{K}}$'$n$ ($n$ forward) would be the same, because both substructures are close to $(M, \,q) \approx (m_K + 2 m_p, ~{\rm{\it lower \,limit}})$.
   
   The 2D plot (Fig.\,\ref{fig:data}a) shows that the event distribution patterns change at $M(Kpp)$. 
   The yield of the lower $M$ region is reduced as a function of $q$, but extends to $q \sim 650$ MeV/$c$.
   The distribution centroid of $M$ does not depend on $q$ within the statistical uncertainty, which allows a bound state interpretation. 
   On the other hand, the distribution centroid of $M$ above $M(Kpp)$ depends on $q$, and the yield vanishes rapidly as a function of $q$.
    The centroid shifts to the heavier $M$ side for the larger $q$, suggesting its non-resonant feature, {\it i.\,e.}\,the propagator's kinetic energy is converted to the relative kinetic energy between $\Lambda$ and $p$, near the lower $q$ boundary.
    Thus, the most natural interpretation would be non-resonant absorption of quasi-free `${\overline{K}}$' by the `$NN$' spectator  (QF$_{{\rm \overline{K}A}}$) due to the final state interaction (FSI). 
     This process can be understood as a part of the quasi-free ${\overline{K}}$ reaction, in which most ${\overline{K}}$s escape from the nucleus, as we published in \cite{hashimoto}.
     Note that there is another change in event distributions at $M(Kpp)$, {\it{i.e.}}, the event density is low close to the $\theta_n = 0$ line below $M(Kpp)$, while it is high above $M(Kpp)$ (this point will be separately discussed in the last section). 
    
   This spectral substructure is in relatively good agreement with that of Sekihara $\!$- $\!$Oset $\!$- $\!$Ramos's spectroscopic function \cite{Sekihara} to account for the observed structure in \cite{sada}.
    Actually, their spectrum has two structures, namely A) a ``$K^- pp$'' pole below the mass threshold $M(Kpp)$ (meson bound state), and B) a QF$_{{\rm \overline{K}A}}$ process above the $M(Kpp)$. 
    Thus, the interpretation of the internal substructures near $M(Kpp)$ is consistent with their theoretical picture.
 
\section{Fitting Procedure}
\begin{figure}[htbp]
 \begin{center}
 \includegraphics[width=10cm]{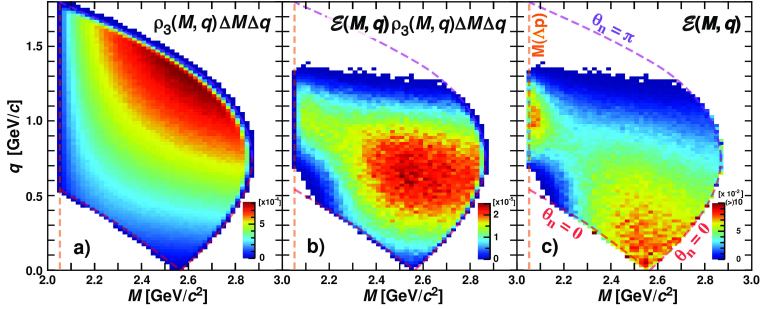}
 \caption{
 \label{fig:PhS-Efficiency}
Simulated spectra of 
a) Lorentz-invariant $\Lambda p n$ phase space $\rho_{3}(M,\,q)$ by taking into account the kaon beam momentum bite, b) $\mathcal{E} (M,\,q) \times \rho_{3}(M,\,q)$, and c) experimental efficiency, $\mathcal{E} (M,\,q)$, evaluated by the bin-by-bin ratio between a) and b).
The unit of $z$-axis (color code) is per one generated events both for a) and b). For c), the ratio is given.
}
 \end{center}
 \end{figure}

\begin{figure}[htbp]
 \begin{center}
 \includegraphics[width=10cm]{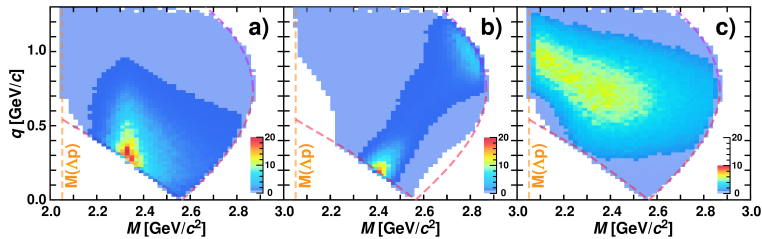}
 \caption{
 \label{fig:f_j}
Individual 2D fit functions of the three physical processes, a) ``$Kpp$'', b) QF$_{{\rm \overline{K}A}}$ and c) $BG$ in the form of $\mathcal{E} (M,\,q) \, \rho_{3}(M,\,q) \,  f_{j} (M,\,q)$ at the best fit parameter set. 
The $z$-axis (color code) is the expected-mean event number to be observed. 
The pale-blue is for the region where the expected number is below one.
The $z$-axis' color code of c) is changed to show its $(M,\,q)$-dependence clearly.
}
 \end{center}
 \end{figure}

    We first describe what we can expect if point-like reactions happen between an incoming $K^-$ and $^3$He, which goes to a $\Lambda p n$ final state. 
    The events must distribute simply according to the $\Lambda p n$ Lorentz-invariant phase space $\rho_{3}(M,\,q)$, as shown in Fig.\,\ref{fig:PhS-Efficiency}a.
    We fully simulated these events based on our experimental setup and analyzed the simulated events by the common analyzer applied to the experimental data.
    The result is shown in Fig.\,\ref{fig:PhS-Efficiency}b, which is simply $\mathcal{E} (M,\,q) \times \rho_{3}(M,\,q)$, where $\mathcal{E} (M,\,q)$ is the experimental efficiency.
    One can evaluate $\mathcal{E} (M,\,q)$ by dividing Fig.\,\ref{fig:PhS-Efficiency}b by Fig.\,\ref{fig:PhS-Efficiency}a bin-by-bin, which is given in Fig.\,\ref{fig:PhS-Efficiency}c.
    As shown in Fig.\,\ref{fig:PhS-Efficiency}c, we have sufficient and smooth experimental efficiency at the region of interest, $M \approx M(Kpp)$ at lower $q$,  based on the careful design of the experimental setup.
    On the other hand, the efficiency is extremely low in the dark-blue to the boundary.
    If we simply apply the acceptance correction, the statistical errors of those bins become huge and very asymmetric.
    This fact makes the acceptance correction of the entire $(M,\,q)$ region unrealistic.
    Therefore, we applied a reverse procedure, {\it i.e.}, we prepared smooth functions $f_{\{j\}}(M,\,q)$ (to account for the $j$-$th$ physical process) and multiplied that with $\mathcal{E} (M,\,q) \times \rho_{3}(M,\,q)$ ({$=$} Fig.\,\ref{fig:PhS-Efficiency}b) bin-by-bin. 
    In this manner, one can reliably estimate how the physics process should be observed in our experimental setup, and this permitted us to calculate the mean-event-number expected in each 2D bin.
      The three introduced model functions (at the best fit parameter set) are shown in Fig.\,\ref{fig:f_j}.

    A very important and striking structure exists below $M(Kpp)$, which could be assigned as the ``$K^-pp$'' signal.
   To make the fitting function as simple as possible, let us examine the event distribution by using the same function as was applied in \cite{sada}, {\it i.e.}, a product of B.W.\,depending only on $M\!,$ and an $S$-wave harmonic-oscillator form-factor depending only on $q$ as:
\begin{eqnarray}
\label{B.W.G.}
\!\!f\!_{\{\!{\rm {\it Kpp}}\!\}} \! = \! 
\frac{ C_{\rm {\it Kpp}} \left( \Gamma\!_{\rm {\it Kpp}} /2 \right)^2 }
{  \left( M - M_{\rm {\it Kpp}} \right)^2 \!\!+ \! \left( \Gamma\!_{\rm {\it Kpp}}  /2  \right)^2 }
\, \exp \!\! \left(\! - \!\left(\!\frac{  q }  { Q_{\rm {\it Kpp}} }\! \right)^{\!\!2}  \right)\!\!.
\end{eqnarray}
where $M_{\rm {\it Kpp}}$ and $\Gamma_{\rm {\it Kpp}}$ are the B.W. pole position and the width,  $Q_{\rm {\it Kpp}}$ is the reaction form-factor parameter, and $C_{\rm {\it Kpp}}$ is the normalization constant, as shown in Fig.\,\ref{fig:f_j}a.

   A model-function of the QF$_{{\rm \overline{K}A}}$ channel, $f_{{\{Q\!F_{\rm \, \overline{K}A} \}}} \,( M, q )$, is introduced as follows. 
   As described, we assume that a `${\overline {K}}$' propagates between the two successive reactions. 
   It consists of 1) $K^-N \rightarrow `{\overline {K}}$'$N$ and 2) non-resonant $`{\overline {K}}$'$ + `NN$'$ \rightarrow \Lambda + p$ in the FSI.
   When the $`{\overline {K}}$'  propagates at momentum $q$ as an on-shell particle in the spectator's rest frame ($\equiv$ laboratory-frame), then the resulting invariant mass $M$ ($\equiv I\!M_{\Lambda p}(`{\overline {K}}\!+\!NN$'$)$) can be given as: 
\begin{eqnarray}
\label{eq:QFA-centroid}
M_{F}(q) = \sqrt{4m_N^2 + m_{K}^2 + 4m_N \sqrt{m_{K}^2 + q^2}},
\end{eqnarray} 
where $m_N$ and $m_{K}$ are the intrinsic mass of the nucleon and the kaon, respectively. 
   The curve originating at $M=M(Kpp)$ in Fig.\,\ref{fig:data}a is the $M_{F}(q)$, which is consistent with the $q$-dependence of QF$_{\rm \, \overline{K}A}$ as shown in the figure.
   Along the line, there are two strong event-concentrations observed at $\theta_n = 0$ ({\it backward $`{\overline {K}}$'}) and $\theta_n = \pi$ ({\it forward $`{\overline {K}}$'}). 
   To account for the distribution, we defined $f_{{\{Q\!F_{\rm \, \overline{K}A} \}}} \, ( M, q )$ as follows.
    For the $q$-direction, we introduced Gaussian and exponential distributions at around the minimum and maximum, respectively, with a constant in between. 
    For the $M$-direction, a Gaussian around $M_{F}(q)$ is applied to account for the spectator's Fermi-motion.
      
   There is another component, widely distributing over the kinematically allowed region of $M$ and $q$, which was previously observed \cite{sada}. 
   In reference \cite{sada}, we simply assumed that the yield of this component was proportional to $\rho_{3}(M,\,q)$. 
   However, with the present much improved statistics, we found that we cannot fit this component with $\rho_{3}(M,\,q)$.
   Compared to $\rho_{3}(M,\,q)$, the yields in the heavier $M$ region and lower $q$ region are much weaker, as shown in the fit curve given in Fig.\,\ref{fig:data}b and c.
   Thus, we {\it phenomenologically} introduced a distribution function, $f_{\{BG\}} (M,\,q)$, similar to Eq.\,\ref{B.W.G.}, but we expanded the $q$-dependent harmonic oscillator term to allow angular momentum up to  $P$-wave, as shown in Fig.\,\ref{fig:f_j}c.

   The data $D(M,\,q)$ can be fitted by using the maximum likelihood method, whose likelihood $lnL_{\{\rm{\it fit}\}}$ is given by a Poisson distribution $P(X=D(M,\,q); \lambda_{D}(M,\,q))$ having mean value $\lambda_{D}(M,\,q)$ at each $(M,\,q)$-bin as:
\begin{eqnarray}
     \label{Poisson}
     lnL_{\{\rm{\it fit}\}} = -\sum_M \,\sum_q \,\ln P(X=D(M,\,q); \lambda_{D}(M,\,q)). 
\end{eqnarray}
   The fitting function $\lambda_{D}(M,\,q)$ is defined as:    
\begin{eqnarray}
     \label{PoissonSum}
\lambda_{D}(M,\,q)  = \mathcal{E} (M,\,q) \, \rho_{3}(M,\,q) \left(  \sum_j \, y_j \, f_{j} (M,\,q) \right),
\end{eqnarray}
where $y_j$ is the yield of the $j$-th physical process, and the first term $\mathcal{E} (M,\,q) \, \rho_{3}(M,\,q)$ is simply Fig.\,\ref{fig:PhS-Efficiency}b.
   
   To examine whether we should introduce more sophisticated model functions, we also studied the following distributions. 
   In the $^3{\rm He}(K^-,\Lambda p)n$ reaction followed by $\Lambda \rightarrow p \pi^-$ decay, there are five kinematically independent observables in total.
   The remaining three kinematical parameters, independent of $M$ and $q$, define the decay kinematics of $``K^-pp" \rightarrow \Lambda p$ and the $\Lambda \rightarrow p \pi^-$ decay asymmetry. 
    Thus, these parameters are sensitive to $J^P$ of the reaction channels.
   For the ``$K^- pp$'' signal, we analyzed events in the window $M = 2.28 \sim 2.38$ GeV/$c^2$ where the major part of the component is located, and  $q =  350 \sim 650$ MeV/$c$ where no severe interference is expected with $f_{{\{QF_{\rm \, \overline{K}A} \}}}$.
      The angular distributions are fairly flat for any of the three kinematical parameters.
   Therefore, the angular distribution is consistent with $S$-wave. 
   Thus, there is no specific reason to introduce any sophisticated terms in addition to Eq.\,\ref{B.W.G.}.
   In fact, a flat distribution is naturally expected if the pole's quantum-number is $J^P = 0^-$. 
   We also analyzed the angular distributions for $f_{{\{QF_{\rm \, \overline{K}A} \}}}$ and $f_{\{BG\}}$.  
   However, again we found no specific reason to introduce further terms.
 
\begin{figure}[htbp]
 \begin{center}
 \includegraphics[width=8cm]{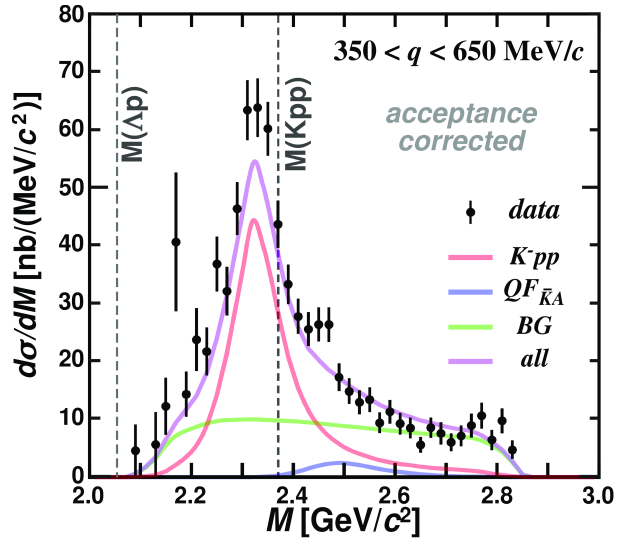}
 \caption{
 \label{fig:q-select}
$\Lambda p$ invariant mass spectrum for $\Lambda  p n$ final state produced in the momentum transfer window of $350 < q < 650$ MeV/$c$. 
   The efficiency $\mathcal{E}(M,\,q)$ was corrected based on the simulation before the $q$ integration of the data.
   Each fitted physical process, which is efficiency corrected and integrated over the $q$-window after the fit, is also given.  
}
 \end{center}
 \end{figure} 

   We haven't considered the interference terms between the three physical processes as given in Eq.\,\ref{PoissonSum}, to avoid over fitting of our statistically limited data.
   Instead, we applied a peak fitting window to reduce the interference effect on our fit result by the following procedures.
   We conducted i) {\it the peak fit}, where $f_{{\{Q\!F_{\rm \, \overline{K}A} \}}} \,( M, q )$ is fitted by fixing all the parameters of $f_{{\{Q\!F_{\rm \, \overline{K}A} \}}} \,( M, q )$ and $f_{\{BG\}} (M,\,q)$ within the $q$-window where no severe interference with QF$_{{\rm \overline{K}A}}$ is expected.
   We then iterated this procedure together with procedure ii) {\it a global fit to evaluate} $f_{{\{Q\!F_{\rm \, \overline{K}A} \}}} \,( M, q )$ {\it and} $f_{\{BG\}} (M,\,q)$ (by fixing parameters in $f\!_{\{\!{\rm {\it Kpp}}\!\}}$ except for the peak yield $C_{\rm {\it Kpp}}$), until procedures i) and ii) converged.

     To exhibit this ``$K^- pp$'' candidate and to present the $M$ spectrum free from experimental acceptance, we plotted the spectrum by correcting our detector efficiency for the events in the momentum transfer window of $350\!<\!q\!<\!650$ MeV/$c$ where mostly $\mathcal{E}(M,\,q) \! \gg \! 0$, as shown in Fig.\,\ref{fig:q-select}.
     To make fit values insensitive to the acceptance correction procedure, we corrected the acceptance as follows.
     The data $D(M,\,q)$ was divided by $\mathcal{E} (M,\,q)$ bin-by-bin and integrated over $q$ at given $M$.
     We applied the same procedure for the data error taking error-propagation into account.
     For each projected physics process $\rho_{3}  f_{j}$ (plotted as the curved lines in the figure), we integrated over $q$, by replacing the $\mathcal{E} (M,\,q) \, \rho_{3}(M,\,q)$ function (given in Fig.\,\ref{fig:PhS-Efficiency}b) with $\rho_{3}(M,\,q)$ (Fig.\,\ref{fig:PhS-Efficiency}a) to be multiplied by $f_{j} (M,\,q)$, {\it c.f.}, Eq.\,\ref{PoissonSum}. 

      In this window, the yield of other processes is largely suppressed in contrast to ``$K^- pp$''.
      The QF$_{{\rm \overline{K}A}}$ distribution is also clearly separated from the ``$K^- pp$'' peak region, because the QF$_{{\rm \overline{K}A}}$ centroid is kinematically shifted to the heavier side, according to Eq.\,\ref{eq:QFA-centroid}, {\it c.f.}, a comparison of the spectral difference of the QF$_{{\rm \overline{K}A}}$ component insetted in blue curves in Fig.\,\ref{fig:data}b and Fig.\,\ref{fig:q-select}.
      As a result, a distinct peak is observed below $M(Kpp)$. 

\section{Fit Result}

   The $S$-wave parameters obtained were; the mass eigenvalue $M_{\rm {\it Kpp}}  = 2324  \pm 3 \,  (stat.) \,^{+6}_{-3} \,(sys.)$ MeV/$c^2$ ({\it i.\,e.} $B_{\rm {\it Kpp}} \equiv  M(Kpp) - M_{\rm {\it Kpp}} = 47 \pm 3 \,  (stat.) \,^{+3}_{-6} \,(sys.)$ MeV), the width $\Gamma_{\rm {\it Kpp}} = 115 \pm 7 \,  (stat.) \,^{+10}_{-20} \,(sys.)$ MeV, and the reaction form-factor parameter $Q_{\rm {\it Kpp}} = 381 \pm 14 \, (stat.)\,^{+57}_{-0} \,(sys.)$ MeV/$c$. 
   The $q$-integrated ``$K^- pp$'' formation yield below the threshold going to the $\Lambda p$ decay channel is evaluated to be $\sigma_{\rm {\it Kpp}} \cdot Br_{\Lambda p} = 7.2 \pm 0.3 \, (stat.) \,^{+0.6}_{-1.0} \,(sys.)$ $\mu$b (for $M<M(Kpp)$).
   For the complete integration over all $q$ and $M$, the cross-section becomes  $\sigma_{\rm {\it Kpp}}^{tot} \cdot Br_{\Lambda p} = 11.8 \pm 0.4 \, (stat.)\,^{+0.2}_{-1.7} \,(sys.)$ $\mu$b.
 
   We evaluated the systematic errors caused by the spectrometer magnetic field strength calibrated by invariant masses of $\Lambda$ and $K^0$ decay, binning effect of the spectrum, and the contamination effects of the other final states ($\Sigma^0 pn$ and $\Sigma^-pp$) to the $\Lambda p n$ event selection.
   To be conservative, the effects to the fit values are added linearly.
   More detailed analysis will be given in a forthcoming full paper.
   
   The $B_{\rm {\it Kpp}}$ $\sim 50$ MeV is much deeper than reported in our first publication since the assumption of a single pole structure was invalid. 
   It is also much deeper than chiral-symmetry-based theoretical predictions.
   The $\Gamma_{\rm {\it Kpp}} \sim$ 110 MeV is rather wide, meaning very absorptive.
   On the other hand, it should be similar to that of $\Lambda(1405) \rightarrow \Sigma\pi$, if ``$K^- pp$'' decays like `$\Lambda(1405)$'$ + $`$p$'$\rightarrow \Sigma \pi p$.
   Thus, the observed large width indicates that the non-mesonic $YN$ channels would be the major decay mode of the ``$K^- pp$''.   
    Interestingly, the observed $Q_{\rm {\it Kpp}} \sim$ 400 MeV/$c$ is very large.
    The large $Q_{\rm {\it Kpp}}$ value implies the formation of a very compact ($\,\sim0.5\,$fm) system referring to $\hbar\sim$\,200\,MeV/$c\,\cdot\,$fm. 
   The compactness of the system is also supported by the large $B_{\rm {\it Kpp}}$. 
   However, the present $Q_{\rm {\it Kpp}}$ can be strongly affected by the primary ${\overline{K}}N \rightarrow {\overline{K}}N$ reaction in the formation process, so one needs more study to evaluate the static form-factor parameter of ``$K^- pp$'' to deduce its size (or nuclear density) more quantitatively.

\section{Discussion and Conclusion}

   We have demonstrated the existence of a peak structure in ${\rm{\it IM}}_{\!\Lambda p}$ below $M(Kpp)$, which can be  kinematically separated very clearly from QF$_{{\rm \overline{K}A}}$ by selecting the momentum transfer window of $350 < q < 650$ MeV/$c$.
    As shown in Fig.\,\ref{fig:data}a, the ``$K^- pp$'' distribution yield reduces near $\theta_n = 0$ as a function of $q$, and it is $\sim$ proportional to the phase space volume defined by Jacobian ({\it c.f.}, Fig.\,\ref{fig:PhS-Efficiency}a (or b)). 
   This is naturally expected if the $S$-wave harmonic-oscillator form-factor given in Eq.\,\ref{B.W.G.} is valid.
    On the other hand, the QF$_{{\rm \overline{K}A}}$ distribution is highly concentrated at $\theta_n = 0$, where the phase space $\rho_{3}(M,\,q)$ is vanishing. 
   This is consistent with our previous result \cite{hashimoto}, in which no structure was found below $M(Kpp)$ at $\theta_n = 0$, {\it{i.e.}},  the leaking-tail of QF$_{{\rm \overline{K}A}}$ into the bound region hides the structure below $M(Kpp)$ at $\theta_n = 0$.

   The present $\Lambda p n$ final state is the simplest channel for $K^-$ interacting with $^3$He.
   In this final state, the ``{\it kinematical anomaly}" is only seen in ${\rm{\it IM}}_{\!\Lambda p}$ having an angular distribution consistent with $S$-wave.       
   Thus, there is no reasonable explanation as to why a peak structure could be formed below $M(Kpp)$ other than ``$K^- pp$''. 
   However, one may wonder whether a spurious bump near $M(Kpp)$ might be formed from some intermediate state converging (or converting) to a $\Lambda p n$ final state in the FSI. 
   
   Here we discuss possible candidates for such an intermediate state.
   Energetically, the possible intermediate states could be `$\Lambda \!+\!p$', `$\Sigma \!+\!N$' and `$\Lambda(1405) \!+\!N$' below `$K^- \!\!+\!p\!+\!p$', which has an $s$-quark and two baryons (`$\Sigma(1385) \!+\!N$' is excluded because it requires $P$-wave).  
   In other words, a $Y^{(*)}$ (baryon with an $s$-quark) could be generated by the primary 2NA reaction, and the $Y^{(*)}$ could make a successive conversion reaction with another spectator nucleon, to form a $\Lambda p n$ final state due to the FSI. 
   Similar to Eq.\,\ref{eq:QFA-centroid}, the ${\rm{\it IM}}_{\!\Lambda p}$ of these channels can be given as:
\begin{eqnarray}
\label{eq:YN-centroid}
   {\rm{\it IM}}_{\!\Lambda p}\left(\,{\mbox{`$Y^{\!(*)}\!+\!N$'}}\,\right) \approx \sqrt{ m_N^2 + m_{{Y}^{\!(*)}}^2 + 2 m_N \sqrt{ m_{{Y}^{(*)}}^2 + q^2} }.
\end{eqnarray}
   First of all, observed ``$K^- pp$'' event concentration does not have the $q$-dependence required by Eq.\,\ref{eq:YN-centroid}.
    Moreover, the ${\rm{\it IM}}_{\!\Lambda p}$ of `$\Lambda \!+\!p$' ($ \sim$2100), `$\Sigma \!+\!N$' ($ \sim$2175 MeV/$c^2$) channels are much too small at the kinematical boundary of $q\sim 500$ MeV/$c$.
    The ${\rm{\it IM}}_{\!\Lambda p}$ of `$\Lambda(1405) \!+\! N$' is $ \sim 2371$ MeV/$c^2$ at the observed average $q$ distribution of $q\sim 450$ MeV/$c$  (assuming $\Lambda(1405)$ mass = 1405.1 MeV/$c^2$ (PDG \cite{Tanabashi:2018oca})).
   Thus the error in the difference from $M_{\rm {\it Kpp}}$ ($\sim 2324$ MeV/$c^2$) is as large as five standard deviations. 
    A direct $\Lambda p$ formation due to 2NA ($K^- + $`$pp$'$\rightarrow \Lambda + p$) could be possible. 
    In this reaction, kaon momentum is 1 GeV/$c$ and the resulting $\Lambda p$ invariant mass $M$ calculated from Eq.\ref{eq:QFA-centroid} is 2.8 GeV/$c^2$.
    In fact, an event concentration is observed at $(M\!c^2, \,qc) \sim (2.8, \,1.0)$ GeV as shown in Fig.\,\ref{fig:data}, but it is well separated from the region of interest. 
    Therefore, none of these can be valid candidates. 
    More complicated channels are even less likely to form a kinematical anomaly at the specific energy near $M(Kpp)$.

The ``$K^- pp$'' signal is significantly above the $M(\Lambda p)$ threshold, so it is unreasonable to explain it as a $\Lambda$-hypernucleus.  
   One may still wonder if the ``$K^- pp$'' signal could be due to the $\Lambda(1405)$ - proton hypernucleus ($_{\Lambda(1405)}^{~~~~~~~2}\rm{H}$), so that the meson (or constituent anti-quark) {\it degree-of-freedom} is already quenched in the system. 
   However, this is not consistent with present data. 
   In the $q$ distribution, the ``$K^- pp$'' signal located at lower $q$ extends up $\sim$ 650 MeV/$c$.
   Thus a tightly bound ``$K^- pp$'' is more natural than a less bound $_{\Lambda(1405)}^{~~~~~~~2}\rm{H}$ measured from $M(\Lambda(1405)p)$.
   It would be even less bound if the pole position of $\Lambda(1405)$ is $\sim$ 1425  rather than 1405 MeV/$c^2$ as the chiral-unitary model suggests \cite{Tanabashi:2018oca}.
   In the $M$ distribution, the signal is much wider than that of $\Lambda(1405)$, so the major decay channel should be $YN$ rather than $\pi\Sigma N$, in contrast to $\Lambda(1405)\rightarrow \pi \Sigma $ (100\%). 
   This drastic change of the decay property is also not consistent with $_{\Lambda(1405)}^{~~~~~~~2}\rm{H}$ interpretation.
   Moreover, if $\Lambda(1405)$ is a ``$K^- p$'' bound system, as recently  accepted rather widely, the discrimination of the two interpretations is meaningless from the beginning.
   
   It is more natural to interpret that the ${\overline K}$ in ``$K^- pp$'' is energetically stabilized ($B_{\rm {\it Kpp}}$ $\sim 50$ MeV) compared to that in ``$K^- p$'' ($\equiv\Lambda(1405)$: $B_{\rm {\it Kp}} \approx 5 \sim 25$ MeV), because of the presence of two protons (nucleons) nearby.
   At the same time, the decay width becomes large ($\Gamma_{\rm {\it Kpp}} \sim$ 110 MeV in respect  to $\Gamma_{\rm {\it Kp}} \sim$ 50 MeV), for the same reason. 
   The existence of the QF$_{{\rm \overline{K}A}}$ channel adjacent to ``$K^- pp$''  also supports this interpretation, because if the sub-threshold virtual `${\overline {K}}$' can form a nuclear bound state by capturing spectator nucleons, then it is natural to expect higher-energy virtual `${\overline {K}}$' production in `vacuum' (above $M(Kpp)$), which could be followed by `$K^-$'$+pp\rightarrow \Lambda p$ in FSI.
   Thus, the simplest and natural interpretation is a kaonic nuclear bound state ``$K^- pp$''; a system composed of a $K^-$-meson and two protons with $J^P=0^-$, {\it I.\,E.} a highly excited novel form of nucleus with a kaon, in which {\it the mesonic degree-of-freedom} still holds.

   In summary, the quasi-free virtual `$\overline{K}$' production $K^-$`$N$'$ \rightarrow $`$\overline{K}$'$N$ is the key reaction in the formation reaction, and $M(Kpp)$ is a doorway below which the ``$K^- pp$'' is formed. 
   Na\"{\i}vely speaking, if the energy of the `$\overline{K}$' produced is below its intrinsic mass ($E_K<m_K$), then the ``$K^- pp$'' will be formed. 
   On the other hand, if it is above the intrinsic mass ($E_K>m_K$), then the QF$_{{\rm \overline{K}A}}$ reaction happens (or the kaon escapes from nuclei).
 





\section*{Acknowledgements}
   The authors are grateful to the staff members of J-PARC/KEK for their extensive efforts especially on the stable operation of the facility. 
   We are also grateful to the fruitful discussions with Professors Akinobu Dote, Toru Harada, Takayasu Sekihara, Khin Swe Myint and Yoshinori Akaishi.
   This work is partly supported by MEXT Grants-in-Aid 26800158, 17K05481, 26287057, 24105003, 14102005 and 17070007. 
   Part of this work is supported by the Ministero degli Affari Esteri e della Cooperazione Internazionale, Direzione Generale per la Promozione del Sistema Paese (MAECI), StrangeMatter project.

\bibliography{E15bibfile}

\end{document}